# Optimizing IaC Configurations: a Case Study Using Nature-inspired Computing

ENEKO OSABA*

TECNALIA, Basque Research and Technology Alliance (BRTA), 48160 Derio, Bizkaia, Spain

GORKA BENGURIA

TECNALIA, Basque Research and Technology Alliance (BRTA), 48160 Derio, Bizkaia, Spain

JESUS L. LOBO

TECNALIA, Basque Research and Technology Alliance (BRTA), 48160 Derio, Bizkaia, Spain

JOSU DIAZ-DE-ARCAYA

TECNALIA, Basque Research and Technology Alliance (BRTA), 48160 Derio, Bizkaia, Spain

JUNCAL ALONSO

TECNALIA, Basque Research and Technology Alliance (BRTA), 48160 Derio, Bizkaia, Spain

IÑAKI ETXANIZ

TECNALIA, Basque Research and Technology Alliance (BRTA), 48160 Derio, Bizkaia, Spain

In the last years, one of the fields of artificial intelligence that has been investigated the most is nature-inspired computing. The research done on this specific topic showcases the interest that sparks in researchers and practitioners, who put their focus on this paradigm because of the adaptability and ability of nature-inspired algorithms to reach high-quality outcomes on a wide range of problems. In fact, this kind of methods has been successfully applied to solve real-world problems in heterogeneous fields such as medicine, transportation, industry, or software engineering. Our main objective with this paper is to describe a tool based on nature-inspired computing for solving a specific software engineering problem. The problem faced consists of optimizing Infrastructure as Code deployment configurations. For this reason, the name of the system is *IaC Optimizer Platform*. A prototypical version of the IOP was described in previous works, in which the functionality of this platform was introduced. With this paper, we take a step forward by describing the final release of the IOP, highlighting its main contribution regarding the current state-of-the-art, and justifying the decisions made on its implementation. Also, we contextualize the IOP within the complete platform in which it is embedded, describing how a user can benefit from its use. To do that, we also present and solve a real-world use case.

---

* Corresponding author: eneko.osaba@tecnalia.com.

## 1 INTRODUCTION

One of the fields of artificial intelligence that is being investigated the most at the moment is nature-inspired computing [1]. In a nutshell, nature-inspired computing aims to create intelligent algorithms by analyzing and mimicking how different kinds of natural phenomena behave. The well-known Genetic Algorithm and Ant Colony Optimization are two of the main sources of inspiration for the conceptualization and subsequent establishment of this field.

The abundant research done on nature-inspired computation demonstrates the interest that sparks in researchers and practitioners, who are drawn to this paradigm due to the ability and adaptability of such methods to obtain near-optimal solutions to a wide range of difficult problems. Specifically, one of the main benefits of nature-inspired algorithms is their capacity to effectively tackle both practical and academic problems. As a matter of fact, the consolidation of this field was the result of many years of fruitful research conducted by a thriving and highly active community, as well as a number of subsequent influential investigations that contributed to the establishment of a number of fundamental concepts.

Since their conception, nature-inspired techniques have been utilized to address a wide range of real-world problems that have arisen in heterogeneous fields like medicine, industry, transportation, energy, and software engineering, among many others [2]. In this context, our main objective with this paper is to describe a nature-inspired based platform for addressing a specific software engineering problem. More concretely, the problem dealt with in this manuscript is focused on optimizing Infrastructure as Code (IaC, [3]) deployment configurations on the most appropriate elements that best meet a set of predefined requirements. The name of the detailed system is *IaC Optimizer Platform* (IOP).

A prototypical version of the IOP was described in previous work [4], in which the functionality of this platform was introduced. With this paper, we take a step forward by describing the final release of the IOP, highlighting its main contribution regarding the current state-of-the-art, and justifying the decisions made on its implementation. Besides, we contextualize the IOP within the complete platform in which it is embedded, describing how a user can benefit from its use. To do that, we also present and solve a real-world use case.

Lastly, it is worth mentioning that the IOP is part of a larger framework built under the umbrella of a Horizon 2020 European research project. The primary objective of this project, known as PIACERE (https://piacere-project.eu/), is to implement a system for the creation, distribution, and management of IaC operating on the cloud continuum.

This manuscript is structured as follows: Section 2 offers some background on the topics dealt on this research. Next, Section 3 describes the IOP, highlighting how it can help users in their daily operations. In Section 4, we introduce a real use case, while Section 5 finishes this paper with several conclusions and future work.

## 2 RELATED WORK

For describing the related work of the research presented in this paper, we have divided this section into two different parts. First, in Section 2.1, we outline some interesting studies focused on a similar research topic. After that, in Section 2.2, we justify the use of the selected optimizing framework by highlighting its benefits in comparison to other state-of-the-art frameworks.

### 2.1 SOTA related to PIACERE project

The optimization problem defined in PIACERE entails having a cloud service that needs to be deployed and a list of infrastructural components, which are based on a list contained in a store called *Infrastructural Elements Catalogue*. The objective is to build the best configuration for deploying the service as efficiently as possible, choosing from the catalogue the most appropriate elements. Although that is a relatively new area of research, there are several studies in the literature that address related issues.

The authors of [5] developed an optimization method inspired by the renowned Non-dominated Sorting Genetic Algorithm (NSGA-II, [6]) that aligns specific Big Data application's requirements with the capabilities provided by an infrastructure as a service (IaaS) and the platform installed inside. That work also studies the Pareto-optima among three different objectives: cost, reliability, and net computing capacity. In [7], authors proposed a similar study, with the goal of fulfilling the non-functional requirements of a concrete microservice while meeting the features established by the developers of this service.

These two works are particularly encouraging for the IOP since the synergy among Cloud Computing and Big Data meets in IaC a practical cloud model, allowing researchers to resort to Big Data features externally and regardless of the service provider [8]. Thus, practitioners can purchase IaaS utilization time on demand based on their specific needs. This is a notion comparable to utility billing for energy or other services [9].

On another note, it is interesting to mention that there are few publications in the literature that discuss how to choose the best infrastructure components to deliver a particular service. In [10], for example, an innovative optimization method is proposed for multi-cloud application deployment. The system presented in that study works in streaming, ensuring that the application is always performing in an appropriate way. The research described in [11] presents a platform that optimizes cluster sizing while leveraging a variety of functionalities, including custom cluster resources or task scheduling characteristics.

Also related to the topic addressed in this paper, considerable research has been conducted related to cloud resource management. In [12], for example, a thorough survey is provided on the specific topic of energy efficiency in cloud computing. Authors categorize optimization techniques based on heuristic solvers, as well as dynamic power management methods. Equally interesting is the work introduced in [13], in which a comprehensive review of different resource provisioning platforms is carried out. Regarding resource management strategies, the authors of that study draw attention to different categories, open research questions. and objective functions.

Lastly, regarding cloud resource configuration, the authors of [14] examine how clusters are sized for deployment in the cloud. They provide a tool that uses a parallel simulation-optimization method to examine various cloud configurations in order to reduce deployment expenses while maintaining quality requirements. In [15], authors address the problem of assisting design-time monitoring of cloud applications for determining the best allocation of components onto virtual machines (VM). To do that, performance requirements and economic costs are considered. In that paper, a tool is proposed that aids users in the process of modeling the structure of an application and mapping each element into a virtual machine.

With all this, and considering the background here described, the IOP developed in the context of the PIACERE project provides the following contributions and novelties:

- The IOP is a flexible mono and multi-objective platform, that allows the user to select which objectives must be optimized from a pool of a predefined set of objectives.
- The IOP provides the user with the possibility of choosing a set of non-functional requirements. Thanks to this feature, the user can define the maximum cost of the overall deployment, or the provider of the elements chosen, among many other constraints.
- The IOP is a multi-algorithm approach that resorts to two well-known multi-objective algorithms (NSGA-II and NSGA-III) depending on the specific needs of the user and the optimization problem built.

## 2.2 Multi-objective solving frameworks

As mentioned in the previous subsection, the problem solved by the IOP is a multi-objective one. The efficient solving of this kind of problems is a recurrent task in artificial intelligence, and because of the popularity of this field, a wide variety of multi-objective frameworks have been proposed by the related community. Based on the analysis conducted in [16], we present in Table 1 a brief review of the characteristics of some of the most representative optimization frameworks. As part of the PIACERE project and the IOP design, a deep analysis has been made comparing the features of these platforms in order to determine which framework will be the most effective to use. For each platform represented in Table 1, we show the programming language employed, the objective of the framework (SOO: Single Objective Optimization, MOO: Multi-Objective Optimization), and its current version.

Table 1: Main features of representative multi-objective optimization frameworks. Regarding the algorithms, in bold are the alternatives for which the framework was originally designed. Current versions checked in July 2023.

| Framework | Language | Algorithms | Current Version |
|---|---|---|---|
| **ECJ** | Java | **SOO**/MOO | 27 |
| **HeuristicLab** | C# | **SOO**/MOO | 3.3.16 |
| **jMetal** | Java | SOO/**MOO** | 6.0 |
| **jMetalPy** | Python | SOO/**MOO** | 1.5.7 |
| **MOEAFramework** | Java | SOO/**MOO** | 3.6 |
| **Pagmo** | C++ | **SOO**/MOO | 2.19.0 |
| **Jenetics** | Java | SOO/MOO | 7.1.3 |
| **PlatEMO** | MATLAB | MOO | 4.2 |

In a brief analysis of Table 1, we can detect that the most commonly used programming language is Java, while other alternatives such as Python, C#, MATLAB, and C++ have also been used. In the context of PIACERE, Java language has been considered appropriate because of its computational efficiency. Regarding the kind of optimization problems that frameworks can deal with, we have bolded these alternatives for which each platform is more dedicated in case that admits both SOO and MOO. For the present research, the specialization of a library has been established as a good reason for choosing one platform above the rest.

With all this in mind, and although several other alternatives could have been used for this purpose – such as MOEA or Jenetics – the platform we selected for implementing the IOP is jMetal [17]. These are the main reasons for the choice:
- jMetal works well with both MOO and SOO.
- jMetal offers a wide set of algorithms for working with many-objective problems.
- Algorithms are flexible to configure and modify in jMetal.
- PIACERE problem definition has been shown easier to perform in jMetal compared to other alternatives.

## 3 FUNDAMENTALS OF THE IOP

As mentioned previously in this paper, the optimization problem resolved by the IOP presents the main challenge of building an optimized deployment configuration of the IaC on suitable infrastructure components that best fits the predefined constraints. To do that, the IOP obtains all the information needed to build the optimization problem from an input data file formatted using a modeling language coined DOML [18].

More specifically, the input DOML contains the objectives that the IOP needs to optimize, and also the non-functional requirements that it must respect. On the one hand, the final release of the IOP allows the user to consider three different objectives: *i)* minimize the cost, *ii)* maximize the availability of the chosen elements, and *iii)* maximize the overall performance of the configuration. On the other hand, six non-functional requirements can be deemed by the IOP, which regard the *i)* cost, *ii)* performance, and *iii)* availability of the whole configuration, as well as the *iv)* provider, *v)* region, and *vii)* memory of the elements chosen.

Once the user defines the information related to the objectives and requirements, the IOP executes the matchmaking against the data available in the catalogue, with the objective of finding the most appropriate elements to deploy. To do that, the IOP resorts to two nature-inspired multi-objective metaheuristics: NSGA-II and NSGA-III [19]. The IOP success depends on whether it is able to offer to the user the most optimized deployment configuration of the IaC. To this end, and because of the multi-objective nature of the problem, several solutions are provided by the IOP, which are ranked by one of the objectives introduced by the user. We refer interested readers to [20] for additional information about the problem formulation and algorithmic design of the IOP.

Once the fundamentals of the IOP have been delucidated, it is appropriate to explain how it works within the PIACERE Ecosystem as a whole. That is, what is the role of the IOP in the overall workflow of the platform? In this regard, the IOP is employed in two different phases: *i)* in the first design of the service (in the so-called *design-time phase*), and *ii)* in the redeployment of an already running service (in the coined *run-time phase*). So, how the IOP can contribute to both phases can be better explained as follows:
- *The IOP in the Design-time phase*: in this step, the IOP is an optional feature that the user can opt to call in case an optimized deployment configuration is needed. For this, the user can directly optimize a properly formatted input DOML using the PIACERE IDE developed as part of the project. In this phase, the IOP offers the user a set composed of (at most) five different optimized solutions, ranked by the objective the user has established as priority. Among the group of solutions found, the IOP chooses the best as the active one. To enhance the understanding of the procedure, we show in Fig. 1 the corresponding workflow diagram. In this regard, it should be clarified that the input DOML is provided to the IOP as part of a compressed ZIP file.
- *The IOP in the run-time phase*: when a service is deployed, the PIACERE Ecosystem analyzes and predicts its performance in real-time. This is done thanks to a module coined *self-learning*. As part of this process, a mechanism named as *self-healing* can require the execution of the IOP for the redeployment of the service if it has failed or if its failure has been predicted by the system. In this specific case, the workflow of the IOP is similar to Fig. 1, considering that the optimization is called by the self-healing mechanism instead of being done through the IDE.

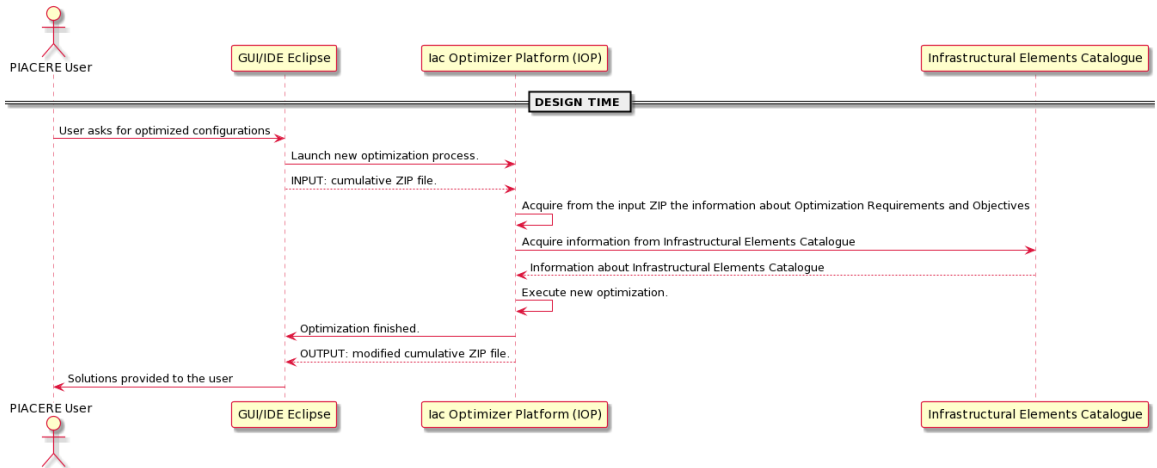

*Figure 1: the workflow diagram of the IOP in the design-time phase.*

Finally, in order to properly understand the overall architecture in which the IOP is involved, it is worth describing in depth the *self-learning* mechanism, whose main role is to analyze the performance of the deployed elements. To do that, the *self-learning* performs a monitoring of some defined parameters (e.g., memory, disk, and CPU), and it is able to predict their values and to deal with some anomalous behaviors that can lead to the failure of the complete service (e.g., anomalies and the concept drift phenomenon). These specific cases would require urgent action by the PIACERE platform, which can involve the redeployment of the service using new infrastructural elements from the catalogue. In other words, *self-learning* is in charge of checking whether elements chosen for the first deployment are running and will run as expected in the short term, and whether they are not suffering from degradation. Going deeper, two different *self-learning* procedures are conducted:
- The *PerformanceSelfLearning*, which is focused on incremental online learning, and which predicts the performance of the elements to guarantee their constant high-level performance.
- The *SecuritySelfLearning*, which resorts to state-of-the-art Natural Language Processing mechanisms to model log streams as a language and capture their normal operating conditions. Anomalous behaviors can also be detected using these models.

Therefore, when a non-desired situation is detected, the *self-learning* component triggers the *self-healing* mechanism, which receives the incidence or the forecast notification. Thus, based on the typology of the notification received, the *self-healing* component identifies a specific mitigation strategy to be applied and proceeds with its execution. As mentioned, one of the mitigation strategies involves the execution of the IOP described in this paper in order to find a new optimized IaC deployment configuration.

## 4 PRACTICAL USE CASE

In this section, a real-world based use case is described in order for the reader to have an idea of how a user can benefit from the use of the IOP. First, as mentioned before, a DOML file should be introduced as input in the IOP with the information related to the optimization properly introduced. This data must be placed in the optimization layer of the DOML. We depict in Fig. 2 an input example. It should be noted that the IOP can deal with much more complex problems. In this article, we present one of reduced complexity for the sake of clarity. Going deeper, in the example depicted in Fig. 2, the user asks the IOP for a deployment configuration composed of a single *Storage* element and a VM. Furthermore, the objectives to optimize are the three available ones: cost, performance, and availability. Lastly, four different requirements are introduced: *i)* the overall cost of the deployment must not exceed 300$, *ii)* the expected availability must be higher than 97%, *iii)* the provider of the elements chosen needs to be Amazon, and *iv)* the memory of the VM should not exceed 1024GB.

```
optimization opt {
    objectives {
        "cost" => min
        "performance" => max
        "availability" => max
    }
    nonfunctional_requirements {
        req1 "cost <= 300" max 300.0 => "cost"
        req2 "availability >= 97%" min 97.0 => "availability"
        req3 "Provider" values "aws" => "provider"
        req4 "max_VM_memory" => "1024"
        req5 "elements" => "VM, Storage"
    }
}
```

*Figure 2: an example of an input DOML for the IOP.*

In addition to this input, and because the IOP seeks an optimized deployment configuration of the IaC on the appropriate infrastructural elements, the user can directly specify in the DOML the VM to which the element selected by the IOP will be mapped. Also, further elements such as VM images, network information, and auto-scale groups can be defined. We show in Fig. 3 an excerpt of the infrastructure layer of the input DOML, in which all this information is introduced.

With all this information, the IOP executes the optimization algorithm, aiming to find the best deployment configurations that meet the requirements introduced by the user. In this case, because three objectives should be optimized, the optimization algorithm chosen is the well-known NSGA-III. Regarding the outcomes, the IOP provides the results in two different formats. On the one hand, the optimization section of the input DOML is extended with general information about the solutions found. In Fig. 4, we depict one example of a solution returned in this format. As it can be seen in that figure, for each solution found, the IOP represents the value of all the objectives defined as well as the identifier of the chosen elements (in this case, `t2.nano` for the VM and `StandardStorage1_Europe` for the *Storage*).

On the other hand, the IOP provides a concretization of the solutions found, which is placed in the concretization layer of the DOML introduced as input. In Fig. 5, where we represent the concretization of the solution shown in Fig. 4, we can see how the details of each chosen element are depicted, which meets the requirements introduced by the user. Several additional aspects should be analyzed in that figure: *i)* the selected VM (t2.nano) is correctly mapped to the OracleDB introduced in the infrastructure layer; *ii)* the network and VM images are also concretized; and *iii)* the auto-scale group is introduced, making reference to the VM found (t2.nano).

It should be considered that the concretization of the solutions is of great importance. In fact, this is the information that the PIACERE Ecosystem employs to finally deploy the complete service.

```
infrastructure abstractInfra {
    // Networks
    net vpc {
        cidr "10.100.0.0/16"
        protocol "TCP/IP"
        subnet subnet1 {
            cidr "10.100.1.0/24"
            connections { subnet1 }
        }
    }
    //VMs
    vm OracleDB {
        os "Ubuntu"
        iface db1 {
            belongs_to subnet1
        }
        sto "1024"
    }
    //VM Image
    vm_image gestaut_vm_image {
        generates gestaut_vm
    }
    // Autoscale group
    autoscale_group gestaut_asg {
        vm gestaut_vm {
            os "Ubuntu"
            iface gestaut_iface {
                belongs_to subnet1
            }
        }
        min 1
        max 1
    }
}
```

*Figure 3: Excerpt of the infrastructure layer with information regarding network, VM, VM image and auto-scale group. This information is used by the IOP for building its solutions.*

```
solution sol1 {
    objectives {
        cost 230.53 euro
        availability 97.5 %
        performance 8.0 metric
    }
    decisions ["StandardStorage1_Europe", "t2.nano"]
}
```

*Figure 4: a possible solution found by the IOP for the input DOML depicted in Fig. 2.*

```
concrete_infrastructure opt_infra1{
    provider aws {
        storage StandardStorage1_Europe {
            properties {
                st_flavor = "StandardStorage1_Europe"
                st_name = "StandardStorage1_Europe"
                st_Availability = 97
                st_Cost_Currency = 130.00
                st_Request_Response_time_Storage_Performance = 4
                st_provider_OU = "aws"
            }
        }
        vm t2_nano{
            properties {
                vm_flavor = "t2_nano"
                vm_name = "t2_nano"
                vm_Availability = 98
                vm_Response_time_Virtual_Machine_Performance = 4
                vm_Memory = 1024
                vm_provider_OU = "aws"
                vm_Cost_Currency = 100.53
            }
            maps OracleDB
        }
        net opt_network_vpc{
            maps vpc
        }
        vm_image concrete_gestaut_vm_image{
            image_name "ami-012e54b30d5c6bc9d"
            maps gestaut_vm_image
        }
        autoscale_group concrete_gestaut_asg{
            properties {
                vm_flavor = "t2_nano"
                vm_name = "t2_nano"
            }
            maps gestaut_asg
        }
    }
}
```

*Figure 5: Concretization of the solution represented in Fig. 4.*

## 5 CONCLUSIONS AND FURTHER WORK

This paper gravitates around a nature-inspired platform for optimizing IaC deployment configurations. The system, named IaC Optimizer Platform (IOP), has been created as part of a European H2020 project. After describing the prototypical version of the tool [4], in this paper we take a step forward by describing the final release of the system. To do that, we have detailed the role of the IOP within the complete PIACERE, and we have shown an example of a real-world use case.

Despite being the final release of the tool, some further work regarding the IOP has been planned. We have the intention of implementing more avant-garde optimization techniques and embedding them into the IOP. We also have the intention of exploring revolutionary computation paradigms, such as quantum computing [21]. Lastly, we have planned to extend the IOP to other application fields related to industry [22], economics [23], or energy [24].

## ACKNOWLEDGMENTS

This research has received funding from the European Union's Horizon 2020 research and innovation programme under grant agreement No: 101000162 (PIACERE project).